\documentclass[]{elsart}
\input psfig.tex
\usepackage[round]{harvard}
\usepackage{harvels}
\begin{document}
\begin{flushright} 
CITA 97-09, DAMTP 97-21
\end{flushright}
\begin{frontmatter}
\title{Correlations Between the Cosmic X-ray \\ and Microwave Backgrounds:\\
Constraints on a Cosmological Constant}
\author[IASHav]{S.P. Boughn}
\author[CITA]{R.G. Crittenden}
\author[DAMTP]{N.G. Turok}
\address[IASHav]{
Institute for Advanced Study, Princeton, NJ  08540 \\
 Department of Astronomy, Haverford College, 
Haverford, PA 19041} 
\address[CITA]{
 CITA, University of Toronto, 
Toronto, Ontario M5S 3H8}
\address[DAMTP]{
DAMTP, University of Cambridge, Cambridge CB3 9EW, UK}
\date{\today}
\begin{abstract}
In universes with significant curvature or cosmological constant,
cosmic microwave background (CMB) anisotropies are created very recently via
the Rees-Sciama or integrated Sachs-Wolfe effects.
This causes the CMB anisotropies to become partially correlated
with the local matter density ($z < 4$).  We examine the prospects of using
the hard (2-10 keV) X-ray background as a probe of the local density and the measured
correlation between the HEAO1 A2 X-ray survey and the 4-year COBE-DMR map to obtain a
constraint on the cosmological constant.  
The 95\% confidence level upper limit on the
cosmological constant is $\Omega_\Lambda \leq 0.5,$
assuming that the observed fluctuations in the X-ray map result entirely from 
large scale structure. 
(This would also imply that the X-rays trace matter with a bias factor of $b_x \simeq 5.6 \ \Omega_m^{0.53}$.)
This bound is weakened considerably if a large portion of the X-ray 
fluctuations arise from Poisson noise from unresolved sources.
For example, if one assumes that the X-ray bias is $b_x = 2.$, 
then the 95\% confidence level upper limit is weaker, $\Omega_\Lambda \leq 0.7$.
More stringent limits should be attainable with
data from the next generation of CMB and X-ray background maps.
\end{abstract}
\end{frontmatter}
\newpage

The recent observations of anisotropies in the microwave background
\cite{ben96a,net97,gun95,sco95}
 have been widely interpreted as 
the result of potential, temperature or velocity
fluctuations at the surface of last scattering, 
originating at very high red shifts ($z > 1000$). 
However, in many cosmological models, a 
significant portion of the anisotropies are produced much more recently.
In critical density cosmologies, 
nonlinear gravitational clustering produces a 
time dependent gravitational potential,
as bound objects turn around and collapse.
This gives rise to CMB anisotropies on small angular scales.
In cosmologies with less than critical density
(e.g. flat universes with a 
large cosmological constant and open universes), time dependence of
the gravitational potential is induced by a change in the expansion 
law of the universe at late times, 
even in the linear perturbation regime.  This gives rise to CMB fluctuations
on larger scales.
The nonlinear effect is usually referred to as the 
Rees-Sciama (RS) effect 
\cite{rs68}, and the linear effect as the 
integrated Sachs-Wolfe (ISW) effect \cite{sw67}.
An earlier 
paper by two of us \cite{ct96} referred to both
as the Rees-Sciama effect.

The recently produced  CMB fluctuations result from time variations 
in the gravitational potential and are correlated with the 
nearby matter density ($z < 4$). 
Observing such correlations could place a strong constraint on 
important cosmological parameters,
such as the matter density, $\Omega_m$, and the 
cosmological constant, $\Omega_\Lambda$. Above all, such correlations 
would
offer a rare opportunity to observe CMB anisotropies 
as they are being produced \cite{ct96}. 

Implementing this, however, requires a probe of the matter density 
at high red shifts.  Possible probes include radio galaxies and
quasars, and a number of large scale surveys of these objects are currently
underway.  In this paper, a map of the hard X-ray background is used
as a measure of the local matter density.
X-rays with energies of order a few keV appear to be produced primarily by 
active galactic nuclei (AGN) \cite{com95} and so should 
reflect the mass distribution on large scales.  
Here, we investigate the cosmological limits which result from  
cross correlating the HEAO1 A2 2-10 keV X-ray map \cite{boldt87} with the four
year COBE DMR map of the cosmic microwave background \cite{ben96a}.

\section{X-ray emission as a tracer of mass} 

In order to use the X-ray background in the type of analysis suggested above two
questions must be answered: ``How well do X-ray sources trace matter?'' and
``What is the redshift distribution of the X-ray emission?''.  The latter of these 
questions will be addressed using a particular model of X-ray sources
\cite{com95} to compute the intensity-redshift distribution,
$d\iota (z)/dz$, i.e. the portion of the intensity of the
X-ray background that arises at redshift $z$.
The results presented in this paper are not
very sensitive to the particular form of $d\iota (z)/dz$ as long as a 
signficant portion of the
intensity arises from redshifts around $\simeq  1$. 
This is the case for most current models of the X-ray background.
How well X-rays trace matter is less well known. 
In this paper we will assume a
simple linear bias, i.e. that the fractional fluctuations in X-ray 
emissivity are linearly 
proportional to the fractional mass density fluctuations, 
$(\delta \epsilon_x /\epsilon_x ) = 
b_x (\delta \rho_m /\rho_m )$, where $b_x$ is the bias factor.  
The linear bias assumption would hold if the X-ray sources were
produced at the high peaks of a Gaussian random field, in which 
case the bias $b_x$ would be approximately 
the height $\nu$ of the
peaks, in units of the standard deviation on the appropriate smoothing
scale.  More generally, it is quite plausible that 
X-ray sources trace the matter density, and the linear bias 
model may be taken as a crude parametrization of this. For example,
if X-ray sources are produced
by local physical processes which are completely independent of the 
long wavelength density perturbations, one would have 
$b_x=1$, whatever the statistics of the density field. 

While the source of the X-ray
background is still very much an area of intense research, it is already 
clear that discrete sources
are the major component.  Deep ROSAT observations have resolved 
$\geq 60\%$
of the soft (0.5 - 2.0 keV) X-ray background into discrete sources
\cite{has93}
and unified AGN (Seyfert galaxies and quasars) models have successfully described both 
the spectrum and the brightness of the hard (2-10 keV) background \cite{com95,mad94,zad93}
 as well as the local volume  
emissivity \cite{miy94b}.  Despite the successes of these models, it 
is quite possible that they will have to be modified in some way (e.g. more detailed 
evolution, refined spectra, a less `unified' AGN model, a new population of sources, etc.) 
as more is learned about X-ray sources.  In particular, correlation and fluctuation analyses
are still not well understood in either the hard or soft bands \cite{miy94b,chen94,car93}.
 However, it seems unlikely that the average flux
distribution, $d\iota (z)/dz$, will be greatly changed by such modifications.  Given that
the ISW effect is not particularly sensitive to the detailed shape of $d\iota (z)/dz$,
we do not believe that such
modifications to the model will have a large effect on the results of this paper.

The estimate of the X-ray bias factor, $b_x$, is more problematic.  Measurements of the
local X-ray bias are still uncertain.  From the dipole anisotropy of the local AGN distribution, 
\citeasnoun{miy94a} estimated that $b_x\Omega_m^{-0.6} \simeq 3.5/f$ where $f$ is fraction of the total 
gravitational acceleration of the Local Group contributed by matter within 45 
$h^{-1}$ Mpc, $\Omega_m $ is the matter density,
and $h$ is the Hubble constant in units of 
$100 \ {\rm km \ s^{-1} \ Mpc^{-1}}$.  For $\Omega_m = 0.3$ and $f \sim 0.5$ this implies $b_x\simeq 3$. 
However, it is not clear if the bias measured in such nearby X-ray sources 
is appropriate on the much larger  scales of interest for the ISW effect.
For example, it is possible that emission from nearby clusters 
might make the local 
bias appear higher.  

In the context of a given theory, where the power spectrum of density fluctuations is 
known and appropriately normalized by the COBE detection of the anisotropies, it is also
possible to place an upper limit on the X-ray bias by observing the  
angular auto-correlation function (ACF) of the X-ray background.
One can compare this to that predicted by the theory assuming a linear, scale-independent bias.
Unfortunately, the large scale ($> 1^\circ$) angular correlations of the X-ray background have
not been accurately measured.  From our own measurements of the ACF of the hard background
we found that, in the context of cold dark matter (CDM) models with a cosmological constant, 
$b_x \simeq 5.6\Omega_m^{0.53}$ (see Table 1).  This should be considered 
as an upper limit since the HEAO beam
size is large ($3^\circ \times 3^\circ$) 
and can itself induce correlations even when viewing randomly 
placed sources.
%and there are possible foregrounds which have not been  < not really a foreground?> 
%subtracted, foremost being fluctuations due to source confusion.
While limits found in this manner are necessarily model dependent, they do 
provide important context for assessing the 
viability of a model.

One might argue by analogy with other populations.  
Ordinary galaxies 
have correlation lengths of order 
$r_o\simeq 5h^{-1}$ Mpc \cite{gp77}
and are thought to be only mildly biased, $b_g\simeq 1$.
This might be considered a reasonable lower limit to the X-ray bias.
The correlation lengths, of
radio galaxies \cite{loa96,sic95}, QSO's 
\cite{ac92}, and groups of galaxies
\cite{bah96} are all on the order of 10 - 18$h^{-1}$ Mpc.  
If, as seems likely, 
hard X-ray sources have similar 
clustering properties to these, then we expect the X-ray bias to be of order
$b_x\simeq 2-3$.  
On the other hand, rich clusters of galaxies are even more strongly
correlated \cite{bah96}, $r_o\simeq 20-25h^{-1} {\rm Mpc}$ and if X-ray sources are instead like these 
it would imply  $b_x > 3.$
It is well known that the X-ray background is highly
correlated with these other objects \cite{ref97,sol96,sol97,miy94b};
however, the correlation length is still uncertain.  
It is clear that the X-ray bias (if indeed linear bias is a good description of the 
distribution of X-ray sources) is not well known.  Therefore, we have chosen to report our 
results with X-ray bias as an unknown parameter.

Finally, as an independent, qualitative check on how X-rays trace mass we have cross-correlated 
the HEAO 2-10 keV map with the appropriately smoothed and binned Greenbank 5 GHz
radio source counts \cite{gre91}.  The correlation coefficient of these two maps 
is signficant, $\langle \delta n \delta \iota\rangle /\delta \iota_{rms} 
\delta n_{rms} =0.2 \pm 0.05$,
indicating that they are tracing correlated populations.  The cross-correlation function 
can also be expressed as 
${\langle \delta n \delta \iota \rangle }\simeq 
{\langle \iota_{R} \rangle + w_{RX}\langle n\rangle \langle \iota_{X} \rangle}$
where $\iota_R $ is the X-ray intensity of the radio sources, $\iota_X $ the intensity of
other X-ray sources in the field, $\langle n\rangle$ is the average number of radio
sources in a resolution element,
and $w_{XR}$ is the cross-correlation coefficient between radio sources and the X-ray sources.
Since the numerical value of $\langle \delta n\delta \iota\rangle /
\langle \iota \rangle$ is 0.03, 
either 3\% of the X-ray background is due to the 5 GHz radio sources or the 
correlation coefficient between radio sources and X-ray sources is on the order of 
$w_{XR} \simeq  10^{-3}$.  
In either case, the implication is that 
X-ray sources are signficantly clustered. 
The Greenbank radio sources are themselves strongly correlated with
correlation lengths of $10 - 18h^{-1}$ Mpc \cite{loa96,sic95}.  When smoothed on the 
HEAO beam scale the radio source correlation coefficient is
$w_{RR} \simeq  2 \times 10^{-3}$.  
These values are comparable to $w_{XX}$ deduced from the X-ray 
auto-correlation function (see Section \ref{sec:acf-d}).  The Greenbank survey is dominated
by moderate redshift ($z\sim 1$) sources \cite{loa96,sic95} which implies (with some
assumption about X-ray source evolution) that most  of contribution to the 
cross-correlation, $\langle \delta n\delta \iota\rangle $, comes from the redshift range 
$0.1 < z < 1$.  This redshift interval is the source of rougly 25\% of the X-ray 
background (see Fig. \ref{fig:FdN}).  Similar results were found cross-correlating the 
X-ray background with the FIRST radio survey \cite{cre96}.  These results cannot 
be made more quantitative without a detailed analysis of the X-ray/radio source 
cross-correlation which is beyond the scope of this paper.

\section{CMB and X-ray Correlations} 

We will focus here on the two point correlations of the 
microwave temperature and X-ray intensity, and the related 
power spectrum of the density fluctuations.  
The density can be written as 
\begin{equation}
\rho({\bf{x}},t) = \bar{\rho}(t)(1 + \delta({\bf{x}},t)), 
\end{equation}
where $\bar{\rho}(t)$ is the average matter density at proper time $t$.  
The power spectrum is defined as the Fourier transform 
of the two point correlation function and is given by, 
$\langle \delta_{\bf k}\delta_{\bf k'}^* \rangle = P_k \delta({\bf k-k'})$ 
where $\delta ({\bf x}, t_0) = \sum_{\bf k} \delta_{\bf k} e^{i
{\bf k\cdot x}} $ and $t_0$ is the present time.  
(Note that we will use ${\bf x}$ and ${\bf k}$ to refer to comoving 
position and wavenumber.)

All sky maps of the X-ray intensity or temperature fluctuations 
can be naturally expanded in terms of spherical harmonics. 
For example, we can write the fluctuation in X-ray intensity as 
\begin{equation}
{\delta \iota({\bf n}) \over \bar{\iota}} = 
\sum_{l,m} a^X_{lm} Y_{lm}(\Omega_{\bf n}),
\end{equation}
where $\delta \iota({\bf n}) = \iota({\bf n}) - \bar{\iota}$ 
and {\bf n} is the unit direction vector.
A similar expression can be written for the temperature 
anisotropy $\delta T ({\bf n})$. 
The expectation value of the moments defines an angular power spectrum 
\begin{equation}
C_l = \langle|a_{lm}|^2\rangle, 
\end{equation}
which is related to the expectation value of the 
autocorrelation function by 
\begin{eqnarray}
C(\theta) & = & \left\langle {\delta \iota({\bf n}) \over \bar{\iota}}
{\delta \iota({\bf n'}) \over \bar{\iota}} \right\rangle \nonumber \\
& = & {1 \over 4\pi} \sum_l (2l+1) C_l P_l(\cos \theta),  
\end{eqnarray}\label{eq:acf}
where $\cos \theta = {\bf n\cdot n'}$ and $P_l(x)$ is a Legendre polynominal. 

\subsection{CMB Anisotropies} 

In the approximation of
instantaneous recombination,
the microwave anisotropy in a direction ${\bf n}$ on the sky is
given in Newtonian gauge by  
\begin{equation}
\frac{\delta T}
{T}({\bf{n}})
 = \bigl[{1\over 4} \delta_\gamma   + {\bf v\cdot n} +\Phi \bigr]_{\tau_r }
^{\tau_0 } +2 \int_{\tau_r }^{\tau_0 } d\tau \dot{\Phi} (\tau, {\bf n}
(\tau_0-\tau)).\label{eq:esw}
\end{equation}
The integral is over the conformal time $\tau $, and 
$\tau_r $ and  $\tau_0$ are the times of recombination and the present, 
respectively.
The first term (${1\over 4} \delta_\gamma $) represents density perturbations
of the radiation-baryon fluid, the second term is the Doppler shift 
(${\bf v\cdot n}$), and the third is the Newtonian potential ($\Phi $), 
where all of these are evaluated on the surface of last scattering.
The last term, usually called the Integrated Sachs-Wolfe (ISW) term,
represents the effect of a time varying gravitional potential
along the line of sight. Heuristically, it represents the redshifting
of photons which `climb out' of  a different potential than they
`fell into'. 

The CMB anisotropies are the sum of contributions 
created near the surface of last scattering and those
created recently by the ISW effect. 
Since the two contributions are associated with perturbations 
of very different wavelengths,
they are nearly 
uncorrelated except on the largest angular scales.
The moments can be written as, 
\begin{equation}
 a^T_{l m} = a_{l m}^{ISW} + a_{l m}^{LS}  
\end{equation}
and the fact that they are weakly correlated 
implies 
\begin{equation}
C^T_l \simeq C_l^{ISW} + C_l^{LS}, 
\end{equation}
where 
$C_l^{ISW} = \langle|a_{lm}^{ISW}|^2\rangle$
and $C_l^{LS} = \langle|a_{lm}^{LS}|^2\rangle$.

The ISW term arises when the CMB photons pass through a 
potential fluctuation which is evolving in time, i.e. 
\begin{equation}
a_{l m}^{ISW}  = 
4 \pi i^l \sum_{\bf k} {\delta_{\bf k} \over P_k^{1/2}} 
\int 2 \dot{\Phi}(k,r) j_l(kr) dr Y_{lm}(\Omega_{\bf k}), 
\end{equation}
where $r$ is the conformal distance, $\tau_0 -\tau,$ and 
$j_l(x)$ is a spherical Bessel function.
Squaring this and integrating over the directions of ${\bf k}$, 
we find
\begin{equation}
C_l^{ISW} =  {2 \over \pi} \int k^2dk 
\left|\int 2 \dot{\Phi}(k,r) j_l(kr) dr \right|^2. 
\end{equation}

\subsection{The X-ray Model} 
 
To evaluate the expected X-ray brightness fluctuations, a model is 
needed which describes the redshift distribution of the sources, as well as their luminousities.
We consider the unified AGN model of \citeasnoun{com95}, 
which reproduces source number
counts and the flux and spectrum of the X-ray background in both soft 
and hard X-ray bands.
In this model, AGN are divided into
two luminosity classes, Seyfert galaxies 
and quasars, which range in X-ray luminosity from $10^{42} {\rm ergs \ s^{-1}}$
to $10^{47} {\rm ergs \ s^{-1}}$ (0.3-3.5 keV).  The power-law luminosity function 
has a break at a luminosity of $L_B=6.9 \times 10^{43} {\rm ergs \ s^{-1}}$ where
AGN with luminosities $L < L_B$ are designated Seyferts and those with $L > L_B$
are designated quasars.

The AGN are thought to be surrounded by an absorbing molecular torus, with the 
amount of absorption depending on the line of sight.  
As an approximation, the AGN are divided into five absorption classes characterized by their 
$H$ column density, 
with column densities ranging from zero up to $N_H = 10^{25} {\rm cm}^{-2}.$ 
The number density of each class is proportional to the number
density of unabsorbed AGN, i.e. $n_{S,i} = w_i n_{S,0}$ and $n_{Q,i} = w_i n_{Q,0}$ 
where $n_{S,0}$ ($n_{Q,0}$) is the total number of unabsorbed Seyferts (quasars) and the index $i$ runs 
over absorption class.
The relative density in each class is assumed to be independent of both source luminosity and redshift.  A detailed description of this division and the associated spectra can be found 
in \citeasnoun{com95}.

The average number of AGN in a volume at redshift $z$ 
is $d\bar{N}(z) = n(z) dV$ where $dV = d_A^2(z) d\Omega dl$, 
$d_A(z)$ is the angular diameter distance, $d\Omega$ is the solid angle 
and $dl$ is proper radial distance.  Proper distance can be expressed as
\begin{equation}
dl = dt = {da \over a'}  = {dz \over 1+z}{a \over a'},
\end{equation}
where ``$a$'' is the expansion factor and primes indicate derivatives with 
respect to proper time. 
In the Comastri model, the comoving number density is assumed to be constant, 
so that $n(z) = n_0 (1+z)^3$ where $n_0$ is the total unabsorbed AGN density.  
It can be expressed as, 
$n_0 =  n_{S,0} + n_{Q,0} = \int_{L_{min}}^{L_B} \phi_s (L) dL + \int_{L_B}^{L_{max}} \phi_q (L) dL$,
where $\phi_s $ is the luminosity function of Seyferts and $\phi_q $ the luminosity 
function of quasars.
 
Another key feature of most X-ray models is that the X-ray luminosity evolves with redshift as 
\begin{equation}
L(z) = L(0)\times (1+z)^\beta.
\end{equation}
The Comastri model assumes $\beta = 2.6$ and a cutoff redshift, $z_{cutoff} = 2.25$,
above which the luminosity is assumed to be constant, up to  
maximum redshift of 
$z_{max} = 4$.
There is some evidence the evolution could be even stronger than assumed by Comastri et al., with a 
slightly higher index, $\beta \simeq 3$, and a somewhat lower cutoff redshift, $z_{cutoff} \simeq 1.8$.
A detailed examination of 
these luminosity models is found in \citeasnoun{boy94}.

The flux received from an object at a given redshift is related to its 
luminosity by
\begin{equation}
{\mathcal F}(z,L) = {L(z) \over 4\pi d_L^2}
\end{equation}
 where the luminosity distance
$d_L$ is related to the angular distance by $d_L = d_A (1+z)^2$. 
For band limited detectors this relation becomes 
\begin{equation}
{\mathcal F}(z,L,I) = {f(z,I) L(z) \over 4\pi d_L^2}. 
\end{equation}
where $L(z)$ now represents the
energy emitted in the detector band $[E_1, E_2]$ and $f(z,I)$ is the usual K-correction
\begin{eqnarray} 
f(z,I) &=& {L_{detected} \over L} = {\int_{E_1(1+z)}^{E_2(1+z)} I(E) dE \over 
\int_{E_1}^{E_2} I(E) dE} \nonumber \\
 &=& {(1+z)\int_{E_1}^{E_2} I(E(1+z)) dE \over 
\int_{E_1}^{E_2} I(E) dE}.
\end{eqnarray} 
Since the K-correction depends on $I(E)$, the spectrum of the source, it is a different
function for each absorbtion class.

At a given redshift the flux averaged over the various types of AGN is obtained by 
summing over both luminosity and absorption, i.e., 
\begin{eqnarray}
{\mathcal F}(z) &=& {\int_{L_{min}}^{L_B} \phi_s (L)  dL \sum w_i {\mathcal F}(z,L,I_{Si})
+ \int_{L_B}^{L_{max}} \phi_q (L) dL\sum w_i {\mathcal F}(z,L,I_{Qi}) \over 
\int_0^\infty \phi(L) dL \sum w_i } \nonumber \\
&=& {(1+z)^{\beta} \over 4\pi d_L^2 } \left({\langle L_S \rangle 
\sum_i w_i {f(z,I_{Si})} + \langle L_Q \rangle
\sum_i w_i {f(z,I_{Qi})}} \right)
\end{eqnarray}  
where $\langle L_S \rangle = \int_{L_{min}}^{L_B} \phi_s (L) L dL /n_0\sum w_i$ and 
$\langle L_Q \rangle = \int_{L_B}^{L_{max}} \phi_q (L) L dL/n_0\sum w_i$. We denote the 
term in parentheses as $L_K(z)$, the average K-corrected luminosity per unit volume.
For the absorbed spectra used by \citename{com95}, this factor can be approximated by 
$L_K(z) \propto (1+z)/(1+ 0.15z)$. 

Finally, the ``space weighting function'', i.e. the contribution to the band limited X-ray intensity 
as a function of redshift,  
is
\begin{equation}
 {d\bar{N}(z) \over dz} {\mathcal F}(z) = n_0 (1+z)^3 d_A^2 d\Omega {1 \over 1+z} 
{a \over a'} {(1+z)^{\beta} \over 4\pi d_L^2 } L_K(z). 
\end{equation}
Using the fact that ${a' / a} = H_0 (\Omega_\Lambda + \Omega_m (1+z)^3)^{1/2}$
for the flat cosmologies studied here, and 
simplifying, the weighting function is 
\begin{equation}
{d\bar{N}(z) \over dz} {\mathcal F}(z) = {n_0(1+z)^{\beta-2}L_K(z) d\Omega 
\over 4\pi H_0 (\Omega_\Lambda + \Omega_m (1+z)^3)^{1/2}}.
\end{equation}

\begin{figure}[htbp]
\centerline{\psfig{file=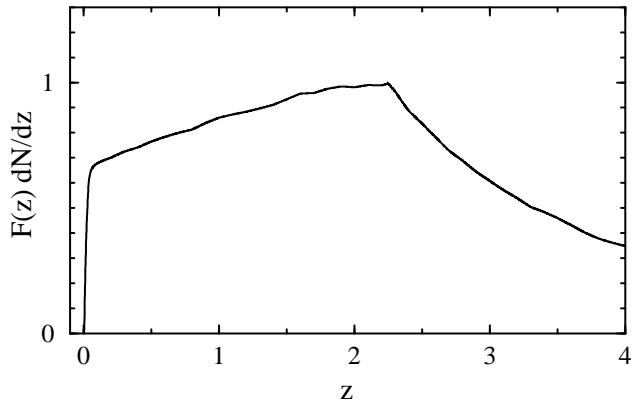,width=3.9in}}
\caption{
The redshift distribution of contributions to the intensity fluctuations in 
the cosmic X-ray background, based on the model of Comastri, et al.  
The cutoff at low redshift results from excluding high flux X-ray sources, 
while the 
cusp at $z=2.25$ results from the change in the luminosity 
evolution of the sources. 
}
\label{fig:FdN}
\end{figure}

Figure \ref{fig:FdN} is a plot of ${d\bar{N}(z) / dz} {\mathcal F}(z)$ in arbitrary units. It is clear from
this figure that much of the flux arises from sources at redshifts of order $z \simeq 1-2$ as
required in order to detect a significant ISW effect.  
The shape of the distribution is fairly flat and is similar to (though it extends to somewhat higher 
redshifts)  
the distribution used in 
\citeasnoun{ct96} which used ${d\bar{N}(z) / dz}$ measured in a flux limited 
survey and assumed that the flux received 
from an object at a given redshift was constant.

It should be noted that for their analysis \citeasnoun{com95} adopted a Hubble constant
of $H_0 = 50 {\rm km s^{-1} Mpc^{-1}}$ and a deceleration parameter of $q_0 = 0$.  The fact
that these differ from the cosmological parameters of the models in this paper is not of
great concern since ultimately we only use the space weighting function which
is directly tied to observations and is, therefore, relatively insensitive to the 
cosmological model.  In any case, as was pointed out previously, the ISW effect 
is not overly sensitive to the space weighting function.

\subsection{X-Ray Correlations}

Given a model for the space weighting function, we then must calculate the 
fluctuations in intensity. 
Let $N({\bf x}, z)$ be the number of AGN in a volume $dV$ at a position 
${\bf x}$ and redshift $z$.  If number density fluctuations are assumed to 
be related to mass density fluctuations by a redshift dependent bias $b_x$, then   
\begin{equation}
 {\delta N({\bf x}, z) \over \bar{N}(z)} = 
{\delta n({\bf x}, z) \over \bar{n}(z)} \equiv 
b_x(z) {\delta \rho({\bf x}, z) \over \bar{\rho}(z)} 
= b_x(z) \delta ({\bf x}, z), 
\end{equation}
where  $\bar{N}(z)$ is the average number of AGN,
and 
\begin{equation}
N({\bf x}, z) = \bar{N}(z) \
(1 +\ b_x(z)\delta ({\bf x}, z)).
\end{equation}
For simplicity, throughout we will assume that the bias is independent of redshift. 
 
The evolution of the mass density fluctuation is 
characterized by, $\delta ({\bf x}, z) = \delta ({\bf x}, 0) D(z).$
For flat, matter dominated cosmologies, perturbations grow 
proportionally to the scale factor, that is, $D(z) = (1+z)^{-1}.$ 
However, for alternative cosmologies such as those with a cosmological 
constant, or with significant curvature, perturbations eventually cease 
to grow. 
 
The intensity of X-rays in a given direction is just the sum of the 
flux from each galaxy, 
\begin{equation}
 \iota({\bf n}) = \int d{N}({\bf x}, z)  {\mathcal F}(z) 
\end{equation}
where ${\mathcal F}(z)$ is the flux we receive from a galaxy 
at redshift $z$ and $r{\bf n} = {\bf x}$. 
Thus, the X-ray intensity fluctuations on the sky are 
\begin{equation}
{\delta \iota({\bf n}) \over \bar{\iota}} = 
{ \int  {\mathcal F}(z) {d\bar{N} \over dz} b_x(z) D(z) \delta ({\bf x}, 0) dz 
\over \int  {\mathcal F} d\bar{N}  }.\label{eq:diota}
\end{equation}

This allows one to solve for the X-ray correlation function. 
By Fourier expanding the present density fluctuation, 
\begin{eqnarray} 
\delta ({\bf x}, 0) & = & \sum_{\bf k} \delta_{\bf k} e^{i
{\bf k\cdot x}}\nonumber \\
& = & \sum_{\bf k}  \delta_{\bf k} 4 \pi \sum_{m,l} i^l j_l(kr)
Y_{lm}(\Omega_{\bf n}) Y_{lm}^*(\Omega_{\bf k}).
\end{eqnarray}
Inserting this into equation (\ref{eq:diota}), one can show that
\begin{eqnarray} 
a_{lm}^X & = & {4 \pi i^l \int  {\mathcal F}(z) {d\bar{N} \over dz} b_x(z) D(z) 
\sum_{\bf k} \delta_{\bf k}  Y_{lm}^*(\Omega_{\bf k}) j_l(kr)
\over \int  {\mathcal F} d\bar{N}  }\nonumber \\ 
& = & 4 \pi i^l \sum_{\bf k} \delta_{\bf k} f_l(k) Y_{lm}^*(\Omega_{\bf k}) 
\end{eqnarray}
where, 
\begin{equation}
f_l(k) = { \int  {\mathcal F}(z) {d\bar{N} \over dz} b_x(z) D(z) j_l(kr) dz
\over \int  {\mathcal F} d\bar{N}  }.
\end{equation}
The argument defines a 
weighting function which determines how deeply in redshift the 
X-rays probe.  
As above, using the expression for the moments and integrating over the 
directions of ${\bf k}$, the X-ray angular power spectrum is
\begin{equation}
C_l^X = {2 \over \pi} \int k^2dk P_k f_l^2(k).   
\end{equation}
If the fluctuations have a Harrison-Zeldovich-Peebles spectrum, $P_k \propto k$, 
then this implies that $C^X_l$ is constant for low multipoles. 
The X-ray autocorrelation function is given by (see Eq. \ref{eq:acf})
\begin{equation}
C_X(\theta)  =  {1 \over 4\pi} \sum_l (2l+1) C_l^X P_l(\cos \theta).
\end{equation}
Note that $C_X(\theta)$ is proportional to $b_x ^2$. This treatment is similar to 
previous calculations of the X-ray background fluctuations \cite{car96,lahav96}. 

In addition to fluctuations from structure, Poisson fluctuations
due to the discreteness of the X-ray sources will also 
contribute to the X-ray auto-correlation function.  
In general, the random placement of nearby objects causes this term to dominate the 
correlation function and the contribution formally diverges if arbitrarily high flux 
objects are considered.  In practice, a high end flux cutoff eliminates the divergence. 
The Poisson contribution to the correlation can be written as 
\begin{equation}
\langle \delta \iota({\bf x}) \delta \iota(0) \rangle_{Poisson} = 
\int d^3x' \bar{n}(z') W(\hat{n},\hat{n}')W(0,\hat{n}') {\mathcal F}^2 (z') 
\end{equation}
where $W(\hat{n},\hat{n}')$ represents the angular beam profile of the detector. 
This contribution to the ACF, which is independent of bias, must be corrected for
in order to compare observations with cosmological models.

\subsection{Cosmological Models and the Power Spectrum}

The models we consider are spatially flat and are primarily 
composed of cold dark matter (CDM) with a cosmological constant. 
We assume the baryon density is constrained by
$\Omega_bh^2 = 0.0125$, though the dependence on baryon density is relatively weak.
The initial fluctuations are assumed to be adiabatic fluctuations with a 
scale invariant Harrison-Zeldovich-Peebles spectrum. 
The present power spectrum is determined from the transfer function and 
to simplify the analysis, we have focused on models which have transfer functions with
the same shape, 
as parameterized by $\Gamma = \Omega_mh{\rm e}^{-(\Omega_b + \Omega_b/\Omega_m)} = 0.25$.
(See, for example, \citeasnoun{ef95} and references therein.) 

We normalize the models using the variance in the CMB on $10^{\circ}$, as measured by 
COBE DMR \cite{ban97}. 
From the growth of perturbations, the power spectrum scales approximately as
$\Omega^{-1.54}_m$\cite{ebw}.  
However,  at large $\Omega_\Lambda$, this dependence is softened due to the 
late time ISW effect.
While a more complete analysis of the model normalization can be 
made using the full COBE data \cite{sug,gor95,wb95,wb96}, using
the $10^{\circ}$ variance is adequate for the present treatment. 
This normalization differs from previous calculations of the X-ray fluctuations 
which normalized with respect to $\sigma_8$, the standard deviation of the mass in a 
sphere of radius 8 $h^{-1} \rm Mpc$\cite{car96}. Roughly, the COBE normalization 
corresponds to a  normalization of $\sigma_8 = 0.7 \ \Omega_m^{-0.6}$ for the 
$\Gamma = 0.25$ models we consider here.

\begin{figure}[htbp]
\centerline{\psfig{file=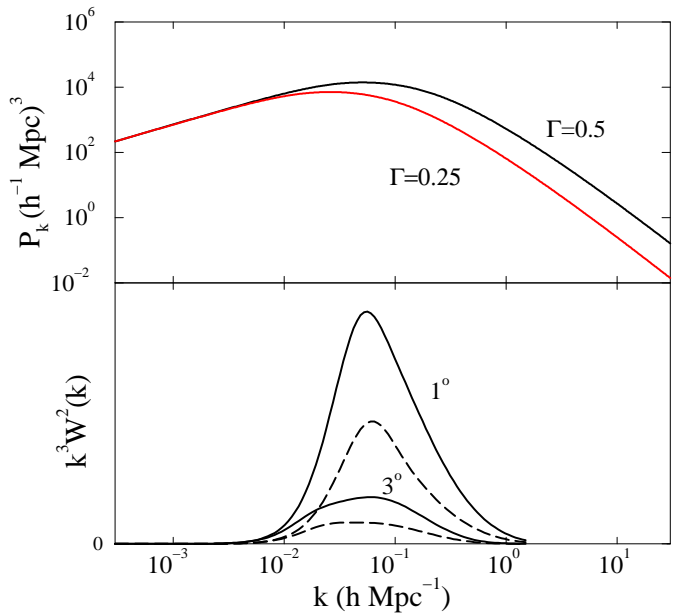,width=3.9in}}
\caption{
The top figure shows the power spectrum for density fluctuations for 
two different values of $\Gamma$.  
The figure below shows where the zero lag X-ray correlation arises 
as a function of wavenumber for different resolutions. 
The X-ray variance, $(\delta X/X)^2$, is proportional to the area underneath the curves. 
Dashed lines represent an $h = 0.32, \Omega_m = 1$ model, while the 
solid lines $h = 0.56, \Omega_m = 0.5$ model.  
}
\label{fig:pk}
\end{figure}

Since we are interested in the X-ray bias, it is important to understand 
what part of the power spectrum is probed by the fluctuations in the X-ray background.
A rough estimate can be made from the flux cut of the survey
and its angular resolution. 
For the given flux cut, $1 \times 10^{-11} {\rm ergs\ cm^{-2}\ s^{-1}}$,
an AGN with mean luminousity $L \simeq 7 \times 10^{43} h^{2}{\rm ergs\ s^{-1}}$ 
would be removed from consideration if it were closer than 
$120 h^{-1}{\rm\ Mpc},$  corresponding to a redshift of $z = 0.04.$ 
Obviously, objects dimmer or brighter than the mean luminousity 
could be closer or further away, but this distance should be 
a reasonable approximation. 
Given a resolution of approximately $3^{\circ}$, this implies that 
the correlation is sensitive to structures on scales greater than 
$6 \ h^{-1}{\rm Mpc}$ or so.  

For a given theoretical model, this can be made more rigorous by 
considering which wavenumbers contribute to the correlation. 
Assuming a Gaussian window function, the smoothed correlation function 
can be written as, 
\begin{equation}
C_X(\theta)  =  {1 \over 4\pi} \sum_l (2l+1) C_l^X P_l(\cos \theta)
e^{-\ell(\ell+1)\theta_B^2/8\log 2}, 
\end{equation}
where $\theta_B$ denotes the FWHM (full width half max) resolution.
From the above expression for $C_l$, we can define 
\begin{equation}
W^2(\theta_B, k) = {1 \over 2\pi^2}
\sum_l (2l+1) P_k f_l^2(k) e^{-\ell(\ell+1)\theta_B^2/8\log 2}.
\end{equation}
(This is analogous to $W_T^2$ defined by \citeasnoun{be87}.) 
With this, $k^3 W^2(\theta_B, k)$ then represents the contribution to the X-ray variance 
for a given resolution from each log interval of wavenumber. 

Figure \ref{fig:pk} shows the contributions versus wavenumber for two 
sample models.  One can see that the correlations arise 
due to modes between $0.01-0.5 {\rm hMpc^{-1}}$, near the peak of the 
power spectrum.  
Thus it spans the gap between those sampled in large scale 
structure surveys and those probed by CMB measurements. 
There are two primary effects which determine the inferred X-ray bias: 
the normalization of the spectrum and the evolution of perturbations, and   
these effects partially offset each other. 
%Large cosmological constant models have less evolution and 
The X-ray bias scales as $\Omega_m^{0.53}$ and $\sigma_X^2$, the area under the curve, 
scales as $\Omega_m^{-1.06}$.  

\subsection{Predictions for the Cross Correlation} \label{sec:ccf}

\begin{figure}[htbp]
\centerline{\psfig{file=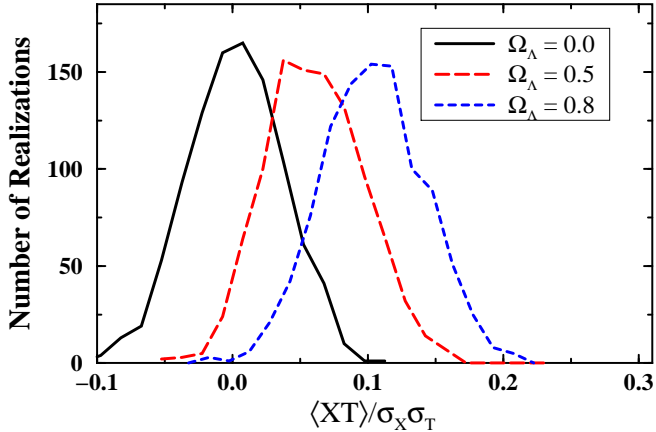,width=3.9in}}
\caption{
Distribution of $\langle XT(0)\rangle$ for 1000 realizations with different values of 
$\Omega_\Lambda$. 
}
\label{fig:hist}
\end{figure}

Only the anisotropies created recently are correlated with the X-ray 
fluctuations described above.  Since that is the case, anisotropies produced 
at last scattering can be ignored and only the ISW contribution to the 
cross correlation need be considered. 
The cross correlation 
between the X-ray and the microwave backgrounds is defined as 
\begin{eqnarray}
C^{XT}(\theta) & = & \left\langle {\delta \iota({\bf n}) \over \bar{\iota}}
{\delta T({\bf n'}) \over T} \right\rangle \nonumber\\
& = & {1 \over 4\pi} \sum_l (2l+1) C_l^{XT} P_l(\cos \theta).
\end{eqnarray} 
Following the same procedures used above, these coefficients can be shown to be  
\begin{eqnarray}
C^{XT}_l & = & \langle a^X_{am} a^{ISW*}_{lm}\rangle\nonumber \\
& = & {2 \over \pi} \int k^2dk f_l(k) P_k^{1/2}
\int 2 \dot{\Phi}(k,r) j_l(kr) dr.  \label{eq:cxt}
\end{eqnarray} 
Because of the statistical nature of the background anisotropies, the correlations
expressed in Equation \ref{eq:cxt} represent ensemble averages.  A particular realization, 
$C_{real}(\theta )$, of $C(\theta )$ can be easily evaluted.  
Assuming each $a_{lm}$ is gaussian distributed with variance $C_l$, then
\begin{eqnarray}
a_{lm}^T & = & \sqrt{C_l^T} \xi_{1,m}\nonumber \\
a_{lm}^X & = & {C_l^{XT} \over \sqrt{C_l^T}} \xi_{1,m}  + 
\left(C_l^X  - {[C_l^{XT}]^2 \over C_l^{T}}\right)^{1/2} \xi_{2,m}
\end{eqnarray}
where $\xi_{i,m}$ are numbers randomly chosen from independent gaussian 
distributions with $\sigma = 1$.  A particular realization is, then,
\begin{eqnarray}
C^{XT}_{real}(\theta) = {1 \over 4\pi} \sum_l (2l+1) R_l^{XT} P_l(\cos \theta)
\end{eqnarray}
where $R_l^{XT} = \sum_m a_{lm}^T a_{lm}^X /(2l+1).$
The distributions of $C^{XT}_{real}(0)$ for 1000 realizations of several different cosmological
constant models are represented in Figure \ref{fig:hist}.  

The spreads of the distributions reflect chance alignments of regions of
enhanced (depleted) X-ray intensity with regions of enhanced (depleted) CMB
intensity.  The latter are due primarily to the CMB anisotropies produced at
last scattering and are, therefore, uncorrelated with X-ray emission.  These
accidental correlations can be thought of as a form of ``cosmic variance''.
The curves in Fig. 3 were computed assuming full sky maps. Limited sky coverage
increases the spread.  It should be noted that widths of the distributions are
approximately independent of $\Lambda $.  The effect of the usual cosmic
variance, i.e. the variance of the actual ISW effect due to the statistical
nature of density fluctuations, is small.
The profiles are roughly Gaussian and we treat
this effect as any other source of noise in the measurement.

Note that we have chosen to plot the quantity
$ C^{XT} (\theta ) / \sigma_X \sigma_T $
which is independent of the bias $b_x$ in the theoretical calculations.  
All of the quantities in this expression are
measureable.  $\sigma_T^2 \equiv {C^T(0)}$ has been measured by COBE and a limit on  $C^{XT} (\theta )$ is
reported in this paper.  However, as discussed above, the measurements of $\sigma_X^2 \equiv C^X(0)$ are still
ambiguous.  In order to compare the model predictions in  Fig. \ref{fig:hist} with
observations, we will take $C^X(0)$  to be that predicted by a given cosmologial model
modulo $b_x ^2 $.  The results will then be given 
as a limit on the cosmological constant
in terms of the unknown X-ray bias.

\section{The HEAO1-A2 and COBE Maps}

The HEAO1 A2 experiment was designed to measure surface brightness in the 
$0.1-60 \ {\rm keV}$ X-ray band \cite{boldt87}.  The present analysis is from two medium
energy detectors (MED) with different fields of view ($3^\circ \times 1.5^\circ $ and 
$3^\circ \times   3^\circ $ FWHM) and two high energy detectors (HED3) with the same fields of view.  
Counts from these four detectors were combined and binned in 24,576 $1.3^\circ \times  1.3^\circ $ pixels
in an equatorial quadrilateralized spherical cube projection on the sky
\cite{ws92}.  The combined map has an effective angular resolution 
diameter (FWHM) of $3.3^\circ $ and a spectral resolution of roughly $2-10 \ {\rm keV}$  
\cite{jm89}.  All data used in this analysis were collected during 
the 6-month period beginning on day 322 of 1977.

The dominant feature in the HEAO map is the Galaxy, so all data within $20^\circ $ of the
Galactic plane and within $30^\circ $ of the Galactic center were cut from the map.  In addition,
$10^\circ $ diameter regions around 90 discrete X-ray sources with $2-10 \ {\rm keV}$ fluxes 
larger than $3 \times  10^{-11} {\rm ergs \ s^{-1} \ cm^{-2}}$ \cite{pic82} were removed.  Finally,
the X-ray map itself was searched for weak point ``sources'' that exceeded the nearby 
background by a specified amount and $7^\circ $ diameter regions around these were removed.
Cuts were made at several levels from 4 to 10 times the photon counting noise
resulting in ``cleaned'' maps with sky coverage of from 26\%  to 47\% .  The analysis
that follows is for the 6 $\sigma$ cut which removed sources with fluxes greater than
$1 \times  10^{-11} {\rm ergs \ s^{-1} \ cm^{-2}}$.  The final ``cleaned'' map corresponds to 1/3 of full
sky coverage; however, the results of this paper are largely independent of the level 
of the cuts.  

Even after cleaning, the HEAO map has several components of large-scale systematic structure
which can be corrected for.  If the dipole moment of the CMB is a kinematic effect,
as it has been widely interpreted \cite{ben96a}, then the hard X-ray background 
should possess a similar dipole structure (Compton-Getting effect) with an amplitude of 
$\delta \iota /\iota  = 4.3 \times  10^{-3} $.  The cleaned map was corrected for this effect.
In addition, a linear time drift in detector sensivity \cite{ja93} results in 
large-scale structure of known form.  Finally, the 2-10 keV background shows evidence
of high latitude Galactic emission as well as emission associated with the Superglactic plane
\cite{ja93}.   We modeled these latter three effects as a combination of
a linear time drift, a Galactic secant law, the Haslam 408 MHz Galactic radio map, 
and a simple pancake Supercluster.  This model was linearly regressed to the cleaned
data and subsquently subtracted from the map.  Of the four parameters, the time drift and
secant law were most significant. 
Correcting for these effects significantly reduced
the large scale structure in the X-ray autocorrelation function but had little effect
at small angular scales as expected (see Section \ref{sec:acf-d}).  When the Compton-Getting dipole
was included in the fit (3 additional parameters), the results did not change significantly.

Because of the ecliptic longitude scan pattern of the HEAO satellite, sky coverage and,
therefore, photon shot noise were not uniform.  However, the mean variance of the cleaned, 
corrected map, $1.63 \times  10^{-2}$ TOT counts/sec, is considerably larger
than the mean variance of photon shot noise, $0.67 \times  10^{-2}$ TOT counts/sec, 
where TOT counts/sec is the
standard HEAO1 A2 normalization for the 2 - 10 keV band \cite{ajw94}.
The correlation analysis
of Section \ref{sec:acf-d} demonstrates that the variance in the HEAO map is due to small angular scale 
($\leq 5^\circ $) intensity fluctuations; therefore, the X-ray map is dominated by ``real'' 
(not photon noise) structure.  For this reason, in the correlation analyses that follow, we
chose to weight each pixel equally. 

The CMB map was constructed from the 53 GHz and 90 GHz 4-year COBE DMR maps as obtained
from the National Space Science Data Center \cite{ben96b}.  Each map consists of 6144
$2.6^\circ \times  2.6^\circ $ pixels in an ecliptic quadrilateralized spherical cube projection, 
i.e. half the resolution of the X-ray map. The 31 GHz maps have
considerably larger instrument noise and Galactic contamination and were not used in 
this analysis \cite{ben96a}.   These four temperature maps 
(A and B channels for each frequency) were converted from antenna to thermodynamic temperature
and then combined in a straight average to form the composite CMB map.  This straight average 
map has somewhat larger instrumental noise than a noise weighted average map; however, the 
noise in the crosscorrelation function is dominated by ``cosmic variance'' (see Section \ref{sec:acf-d})
and we felt that a straight average would be more likely to minimized unknown systematic
effects in the composite map.  The same Galatic cut (within $20^\circ $ of the Galactic plane
and within $30^\circ $ of the Galactic center) was applied to the CMB map as to the X-ray map
which results in 64\%  sky coverage.

The only correction made to the cleaned composite CMB map was to fit and remove the dipole
moment. Since the quadrupole moment was included in the analysis of section II-C,  no 
attempt was made to remove it from the data.  The removal of a secant law fit to high 
Galactic latitude emission made no significant difference in the cross-correlation function
(see Section \ref{sec:acf-d}) and so the CMB map was not corrected for Galaxy emission.  In any case,
the analysis of this paper concerns the  shape of the cross-correlation function
at relatively small angular scales ($\leq 20^\circ $) and is not overly sensitive to the
presence of large scale structure in the map.

As with HEAO, the COBE satellite sky coverage was not uniform; therefore, instrument noise per
pixel is also not uniform over the sky.  In this case the mean variance in the cleaned,
corrected CMB map is  $5.0 \times {\rm 10^{-3} mK^2}$ whereas the mean variance of instrument noise
is $4.1 \times  10^{-3} {\rm mK^2}$, i.e. at the resolution of the map, instrument noise dominates
real structure.  However, when smoothed in $10^\circ $ bins, 
instrument noise variance becomes $3.4 \times  10^{-4} {\rm mK^2}$ compared to the
$9.0 \times  10^{-4} {\rm mK^2}$ fluctuations of the CMB \cite{ban96}.  Therefore, in 
the correlation analyses that follow, we also chose to weight each CMB pixel equally.

\subsection{The X-ray Auto-Correlation Function} \label{sec:acf-d}

The intensity auto-correlation (ACF) is calculated using, 
\begin{equation}
 \langle XX (\theta) \rangle = \sum_{i,j} (\iota_i - \bar{\iota})(\iota_j - \bar{\iota})/N_{\theta},
\end{equation}
where the sum is over pair of pixels separated by $\theta$ and $N_\theta$ is the 
number of such pairs.  Note that the pairs are given equal weight as discussed above.

\begin{figure}[htbp]
\centerline{\psfig{file=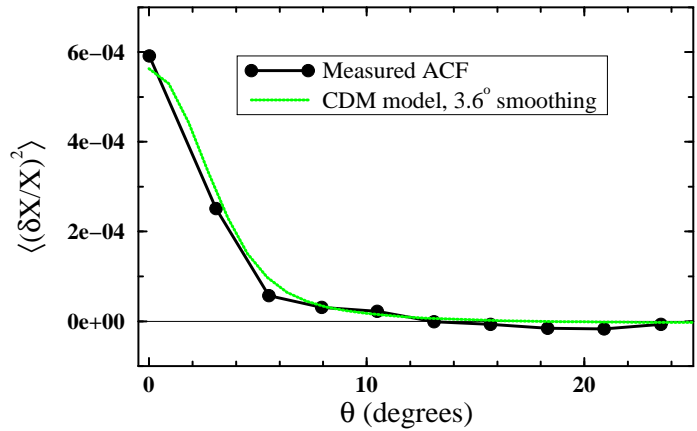,width=3.9in}}
\caption{
The auto-correlation function for the HEAO1-A1 X-ray map, normalized in terms of the mean intensity.   
Shown for comparison is the theoretical prediction for a COBE normalized CDM model with 
an effective smoothing of $3.6^\circ$ FWHM. 
}
\label{fig:acf}
\end{figure}

The results for the cleaned X-ray map are shown in Fig. \ref{fig:acf}.   
The value at $\theta  = 0$ has been corrected for contribution of uncorrelated 
photon shot noise.  To check for the contribution to the ACF of weak X-ray sources
in nearby galaxies, $4^\circ $ diameter holes around all (2367) galaxies in the 
Tully Atlas of Nearby Galaxies \cite{tul88} were cut from the map.  These cuts did not
significantly change the ACF.

Taking into account the finite angular resolution of the HEAO map, the amplitude of the
ACF is somewhat smaller than that found by \citeasnoun{sol96} for the ROSAT 
$0.5-2 \ {\rm keV}$.  However, it should be noted that the Soltan et al. results are inconsistent
with other measurements \cite{chen94,car93} leading them to
hypothesize diffuse emission from large (20 Mpc) clouds of gas.

It should be emphasized that the ACF in Fig. \ref{fig:acf} has not been corrected for Poisson 
noise due to unresolved sources.  The model of \citeasnoun{com95} indicates that
source confusion is substantially less that the measured value while the source counts
of \citeasnoun{pic82} and Butcher (see \cite{car93}) imply a substantial
amount of source confusion.  On the other hand, poor
angular resolution results in shape of the ACF being dominated by the beam profile.  For
these reasons we have not chosen to estimate the correction for source confusion and the
data in Fig. \ref{fig:acf} should be considered an upper limit to the ACF due to clustering of
X-ray sources.

Fig. \ref{fig:acf} also shows the theoretical ACF arising from a CDM power spectrum
with $\Lambda = 0.5$, ($\Gamma = 0.25$), a bias of $b_x  = 3.9$, and smoothed by a
$3.6^\circ$ gaussian beam.  Recall that the height of the theoretical ACF is proportional
to $b_x ^2$.  The sources
were assumed to be distributed as in the model of \citeasnoun{com95} with a flux
cutoff equal to that applied to the real data.  Clearly this bias is a good fit to the 
model; however, to the extent that the observed ACF is contaminated by source confusion, it
should be considered an upper limit for this model.  
Raising $\Gamma$ has the effect of 
adding small scale power, and so would reduce the bias required to fit the X-ray ACF.

\subsection{The HEAO/COBE Cross Correlation} \label{sec:ccf-d}

The cleaned, corrected X-ray map was reprojected onto the sky to match the COBE map, i.e.,
an ecliptic quadrilateralized spherical cube projection with $2.6^\circ \times 2.6^\circ $ pixels.  Let
${\rm w}_i$ be the number of original pixels combined to form one pixel in the new projection.
The value of ${\rm w}_i$ for most pixels was 4; however, because of the transformation to
ecliptic coordinates, ${\rm w}_i$ ranged from 1 to 6.  The terms in the cross-correlation function
(CCF) are weighted with these values.  The cross correlation is then computed as, 
\begin{equation}
 \langle TX(\theta) \rangle = {\sum_{\theta_{ij} = \theta} X_i {\rm w}_i T_j / 
\sum_{\theta_{ij} = \theta} {\rm w}_i}, 
\end{equation}
where the sum is over all pairs with angular separation $\theta$.

\begin{figure}[htbp]
\centerline{\psfig{file=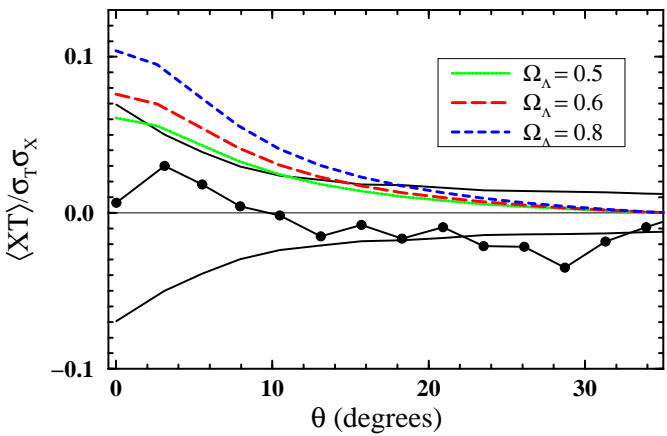,width=3.9in}}
\caption{
The cross correlation between the four year COBE map and the HEAO1-A2 X-ray map, normalized 
by the rms fluctuation of each map.  Also shown are the expected correlations for 
a range of values for the cosmological constant.  (The standard CDM model predicts
no correlation.)  The bold lines show the rms correlations resulting from noise and
cosmic variance.
}
\label{fig:ccf}
\end{figure}

The resulting cross correlation is shown in Fig. \ref{fig:ccf}, where the normalization 
is with repect to the $measured$ rms fluctuations of the two maps.
Also shown are the theoretical predictions for cosmological constant models 
with varying $\Lambda $.  As described previously, the Hubble constant has also been varied
to keep the shape parameter, $\Gamma$, fixed.
Finally, the solid curves represent the $1 \sigma $ uncertainty due to the combination of cosmic
variance, instrument noise, and photon shot noise. 
The spread in the distribution
illustrated in Fig. \ref{fig:hist} did not include the effects of map cleaning, map
correction, and reduced sky coverage.  In addition, there are quite likely additional
fluctuations in the X-ray map that do not appear the X-ray model, e.g.
relatively nearby clusters of galaxies.  Therefore, we chose to compute the
total ``noise'' by a Monte Carlo method wherein the real, cleaned X-ray map was
cross-correlated with an ensemble of 400 randomly generated CMB maps.  These
latter maps consisted of random realizations of instrument noise and cosmic
structure which was constructed so as to have the same ACF profile as the actual
COBE data. All the Monte Carlo CMB maps were cleaned (Galaxy cut) and corrected
(dipole removal) with the same procedure as used on the real map.  The resultant
total noise is correlated on angular scales of about $10^\circ $.  This is due
primarily to the finite angular resolution of the COBE map.  It is clear from
Fig. \ref{fig:ccf} that the data are consistent with the hypothesis that the
two maps are uncorrelated.

The above results assume that fluctuations in the X-ray map are dominated by 
Poisson fluctuations of unresolved sources.
If there is a significant contribution to the X-rays from this Poisson 
noise, then both the inferred X-ray bias and the predictions for the 
theoretical correlation are reduced by a factor of 
$\sigma_{Structure}/(\sigma_{Structure}^2 + \sigma_{Poisson}^2)^{1 \over 2}.$
Therefore, for any particular bias, the model amplitudes should be scaled as
$b_x /5.6\ \Omega_m^{0.53}$.  In any case, it seems unlikely that the bias would
be less than unity.

The upper limit of the CCF at $\theta  = 0$ is consistent with previous correlation
analyses \cite{bj93,ban96} between HEAO1-A2 and temperature anisotropy maps.  
The correlation analysis of the
COBE DMR maps and the ROSAT PSPC All-Sky Survey \cite{kne97} shows somewhat more 
structure than Fig. \ref{fig:acf}.  This is presumably due to a larger Galactic
contriubtion at 1 keV.

\section{Discussion} 

As previously discussed, the observed HEAO/COBE cross-correlation is consistant 
with there being no underlying ISW effect, i.e. no cosmological constant. 
Since the effective errors computed by Monte Carlo analysis (see Section \ref{sec:ccf-d})
are consistent with being derived from a correlated gaussian distribution, 
we can quantify this by performing a maximum likelihood analysis. 
The likelihood is defined by 
\begin{equation}
{\mathcal L} = (2\pi)^{-N/2} [{\rm det} \ R_{ij}]^{-{1 \over 2}}
\exp [-{1 \over 2}R^{-1}_{ij}(y_i - \bar{y}_i)(y_j - \bar{y}_j)], 
\end{equation}
where $R_{ij}$ is the covariance matrix and $\bar{y}_i$ is the theoretical prediction for the ${\rm i^{th}}$ 
observation. 
To a good approximation, the theoretical cross-correlation curves in Fig. \ref{fig:ccf}
have the same shape and only differ in amplitude (which depends only on $\Omega_\Lambda $),
\begin{equation}
\langle TX(\theta) \rangle = A(\Lambda) f(\theta) 
\end{equation}
where $f(0^\circ) = 1$.  
Maximizing the likelihood for this amplitude, one finds, 
\begin{equation}
A = \sum_{ij} R^{-1}_{ij}(f_iy_j + f_jy_i)/2\sum_{ij}R^{-1}_{ij}f_if_j 
\end{equation}
and the variance in this amplitude is given by, 
\begin{equation}
\sigma_A^2 = \sum_{ijkl}R^{-1}_{ij}R^{-1}_{kl}(R_{jl}f_if_k + {\rm 3 \ permutations}\ )
%+ R_{jk}f_if_l + R_{il}f_jf_k
%+ R_{ik}f_jf_l)
/ (2\sum_{ij}R^{-1}_{ij}f_if_j)^2.
\end{equation}
Since ``cosmic variance'' is the dominant uncertainty, 
we calculate the covariance matrix from the correlations of the X-ray map with the 
simulated CMB maps.
Although this implicitly assumes that there are no correlations between the maps, 
it is a good approximation when the ISW effect is small as discussed in Section \ref{sec:ccf}.

The most likely amplitude is found to be 
$A = -0.008 \pm 0.042$, based on the measured correlation for $\theta < 25^\circ$. 
(At larger separations the models predict little correlation.) 
This corresponds to a 95 \% CL upper limit of $A < 0.061$ ($1.65 \sigma$), and a 
98 \% CL upper limit of $A < 0.078$ ($2.05 \sigma$).
The amplitudes, $A$, for different models are summarized in Table 1 along with
the X-ray biases inferred from the X-ray map assuming all of the X-ray fluctuations result from 
large scale structure.  Under this assumption, $\Omega_\Lambda > 0.5$ is just ruled out at the 
95 \% CL. 
As discussed above, if there is a significant 
contribution to the X-rays from Poisson noise, then both the inferred X-ray bias 
and the predictions for the theoretical correlation are reduced by a factor of 
$\sigma_{Structure}/(\sigma_{Structure}^2 + \sigma_{Poisson}^2)^{1 \over 2}.$
Included in the table are the expected amplitudes for a fixed X-ray bias, 
$b_x = 1$ or $b_x = 2.$ 
If $b_x = 2$, $\Omega_\Lambda > 0.7$ is still ruled out at the 95 \% CL. 
In the most pessimistic case, with $b_x = 1$, none of the models can be ruled out. 
%These limits are fairly conservative, as we have focused on $\Gamma = 0.25$ models and 
%larger $\Gamma$ models would have more small scale structure and require smaller biases.

\begin{table} \label{tab:res}
\begin{center}
\begin{tabular}{|c|c|c|c|c|}
\hline
$\ \Omega_\Lambda \ $  & \ Implied $b_x$ \ 
  & $A$, Implied $b_x$ & $A$, $b_x = 2$ & $A$, $b_x = 1$ \\
\hline
$0.0$   & $5.6$  & $0.000$ & $0.000$ & $0.000$  \\
$0.5$   & $3.9$  & $0.061$ & $0.031$ & $0.016$  \\
$0.6$   & $3.4$  & $0.077$ & $0.045$ & $0.023$  \\
$0.7$   & $2.9$  & $0.091$ & $0.063$ & $0.031$  \\
$0.8$   & $2.4$  & $0.104$ & $0.086$ & $0.043$  \\
\hline 
\end{tabular}
\caption{The inferred biases and the expected cross correlation amplitutudes, $\langle XT(0^\circ) \rangle/ 
\sigma_X \sigma_T$ for a range of models with $\Gamma = 0.25$.  The inferred biases and the first column of amplitudes
assume that the X-ray fluctuations result solely from large scale structure.  The other amplitudes assume a 
fixed X-ray bias, $b_x = 1$ or $b_x = 2.$}

\end{center}
\end{table}

The analysis we have presented has been conservative in the
sense that we have only used the X-ray temperature cross
correlation to constrain the models. For nonzero cosmological constant, 
the spectrum $C_l$ of temperature anisotropies shows a 
rise at low $l$ which is absent in the COBE spectrum. 
Inclusion of the latter in our analysis would therefore 
strengthen our bound on $\Omega_\Lambda$.
(For example, see \citeasnoun{ss94}.)  We have chosen not to
do so because the low $l$ rise is a more model-dependent 
effect than the X-ray temperature correlation, it being possible
to remove the former completely by a tilt in the primordial 
power spectrum. Nevertheless, a future analysis of 
a wider class of models should include the $C_l$ 
information as well. 

It is important to search for as many independent constraints 
on a cosmological constant as possible, because 
each is subject to 
different kinds of systematic errors.
At present, the best established limits on $\Lambda$ come from gravitational lensing  
studies \cite{mr93,koch95}.  \citeasnoun{koch95} finds that the lack of observed lensing events implies 
$\Omega_\Lambda < 0.66$ at the 95\% confidence level.  
In addition, recent studies of the deceleration parameter measured in supernovae searches 
have placed an upper limit of $\Omega_\Lambda < 0.51$ at the $2\sigma$ level \cite{per96}.

The limit on the cross-correlation amplitude 
can also be used to constrain open models, where the correlations could be 
much higher for a given $\Omega_m$, if the X-rays probe at high enough redshifts \cite{kam96}. 
The constraints on open models thus are likely to be much more sensitive to assumptions of 
the luminousity evolution of the sources than the cosmological constant constraints.

\section{Conclusions} 

Our primary aim in this paper has been to demonstrate 
how correlations, or the absence of correlations, between 
the microwave background and deep probes of structure can lead to 
constraints on cosmological models. 
The surveys we used were not ideal for this purpose, but even so 
they resulted in a relevant constraint on the cosmological constant.

Surveys with better angular resolution will improve the signal to 
to noise and thus the confidence with which we can rule out a particular cosmology. 
With degree scale resolution in both the temperature and X-ray maps, 
the signal to noise could be improved by 50\%.
For the linear effect which we are considering here, little is gained 
by looking at maps with resolution
on scales smaller than a degree or two. 
The CMB photons do not have time to 
receive a substantial shift from the time varying potential 
when passing through smaller fluctuations.
Since the correlation is restricted to large scales, 
full sky maps are essential to reduce uncertainties due to accidental 
correlations. 

Future satellite missions, most notably MAP and the Planck Surveyor, will 
provide full sky maps with sufficient resolution ($0.3^\circ$ and $0.1^\circ$, respectively)
to be explore this effect fully in the CMB sky.  
Data from these probes should be available in the next 5-10 years. 
The prospects for better data for the X-ray sky in the near future are uncertain. 
The ROSAT survey is sensitive only at lower energies ($<2$ keV) and so is 
likely to be contaminated by galactic and other foregrounds. 
To search for this effect, ideally one needs a full sky survey of the hard XRB (2-10 keV) 
which has degree scale resolution and which is sensitive enough to allow an unambiguous 
subtraction of Poisson fluctuations.  Such a survey would also provide a measurement of 
structure on scales larger than is possible in current optical surveys and would 
fill in the gap in the power spectrum which 
lies between the galactic surveys and the measurements of the microwave background. 
This kind of mission is now being considered (K. Jahoda, private communication). 

One can also consider looking at 
other possible probes of structure at high red shifts.
The formalism for calculating the correlations presented here for X-rays transfers easily 
to other kinds of probes.
Possibilities include 
surveys of radio galaxies or quasars, or other objects which 
probe to large redshifts.  
Galaxy surveys such as the Sloane DSS, which should probe to $z\sim 0.5$, 
might also be sensitive to this effect and have the added benefit of containing 
redshift information for the individual sources.
An advantage of these probes is that their large scale clustering properties
are (or will be) known and the bias implied by a given model can be easily
computed.
However, if the surface density of these objects is too low (e.g. quasars) then
the statistical noise in the cross-correlation function will be large and the
effects discussed in this paper will be unobservable.
 
In this paper we have focused on testing for a cosmological constant, but in fact 
this is a rather indirect means of testing for $\Lambda$. The results we 
quote are based on a fairly specific model, one which has cold dark matter as well as
adiabatic, scale invariant initial fluctuations. 
It is important to realize however that rather than being sensitive merely to 
one specific cosmological model, the effect we have considered occurs in 
any cosmology where the gravitational potential has evolved recently. 
This includes models that are open or where structures arise from cosmological 
defects and we are presently working to understand this constraint in these other contexts.

Whatever the cosmological model, the ISW effect reflects important fundamental 
information relevant to structure formation.
The presence of large, empty voids, the apparent lack of mergers and the 
shapes of clusters of galaxies provide evidence, tentative as yet, that the process of structure 
formation might be slowing down.  
If this is the case, then the decay of the linear gravitational potential 
makes a late time ISW effect inevitable and offers an important  
observation to confront these other clues about how structures formed in the universe.

\begin{ack}
We thank Elihu Boldt and Jim Peebles for useful discussions.  The HEAO1 A2 data and map
generating software were kindly provided to us by Keith Jahoda.  Much of the data analysis was
done at Princeton University and we thank Ed Groth for software and computing support.  This 
work was supported in part by NASA grant \#NAG 5-3015, NSF grant \#PHY-9222952, and the Monell
Foundation.
\end{ack}

\end{document}